\documentclass[%
 reprint,
superscriptaddress,
showpacs,preprintnumbers,
 amsmath,amssymb,
aps,
prl,
]{revtex4-1}

\usepackage{lineno}

\usepackage{comment}

\usepackage{txfonts}
\usepackage{amsfonts}
\usepackage{color}%
\usepackage{graphicx}
\usepackage{dcolumn}
\usepackage{bm}

\begin{document}


\title{
Topological Dirac Nodal Lines and Surface Charges
in fcc  
Alkaline Earth Metals
}

\author{Motoaki Hirayama}
\affiliation{%
 Department of Physics, Tokyo Institute of Technology, Ookayama, Meguro-ku, Tokyo 152-8551, Japan 
}%
\affiliation{%
TIES, Tokyo Institute of Technology, Ookayama, Meguro-ku, Tokyo 152-8551, Japan 
}%

\author{Ryo Okugawa}
\affiliation{%
 Department of Physics, Tokyo Institute of Technology, Ookayama, Meguro-ku, Tokyo 152-8551, Japan 
}%

\author{Takashi Miyake}
\affiliation{%
Research Center for Computational Design of Advanced Functional Materials, AIST, Tsukuba 305-8568, Japan 
}%

\author{Shuichi Murakami$^*$}
\affiliation{%
 Department of Physics, Tokyo Institute of Technology, Ookayama, Meguro-ku, Tokyo 152-8551, Japan 
}%
\affiliation{%
TIES, Tokyo Institute of Technology, Ookayama, Meguro-ku, Tokyo 152-8551, Japan 
}%

\date{\today}

\begin{abstract}
In nodal-line semimetals, 
the gaps close along loops in ${\bf k}$ space,
which are not at high-symmetry points. 
Typical mechanisms 
for the emergence of nodal lines involve mirror symmetry and the $\pi$ Berry phase.
Here, we show via \textit{ab initio} calculations that fcc calcium (Ca), strontium (Sr) and 
ytterbium (Yb) have topological nodal lines with the $\pi$ Berry phase near the Fermi level, when spin-orbit interaction is neglected. In particular, Ca becomes a 
nodal-line semimetal at high pressure.
Owing to nodal lines, the Zak phase 
becomes either $\pi$ or $0$ depending 
on the wavavector ${\bf k}$, and the $\pi$ Zak phase leads to surface polarization charge. 
Carriers eventually screen it, 
leaving 
behind large surface dipoles.
In materials with 
nodal lines, 
both the large surface polarization charge and the emergent drumhead surface states 
enhance Rashba splitting when heavy adatoms are present, as we have shown to occur in Bi/Sr(111) and in Bi/Ag(111).
\end{abstract}


\maketitle


\noindent{
Recent discoveries of topological semimetals have taught us that  the
${\bf k}$-space topological structure of electronic bands plays a vital role 
in a number of materials. This class of topological semimetals includes 
Weyl semimetals \cite{Murakami07b,Wan12}, Dirac semimetals \cite{Wang12,Wang13}, and nodal-line semimetals (NLSs)
\cite{Mullen15,Fang15,Chen15nano,Weng15b,Kim15,Yu15,Xie15,Chan15,Zeng15,Yamakage16,Zhao15}.
In topological semimetals, the conduction and valence bands touch 
each other at some generic points (as in Dirac and Weyl semimetals) or 
along curves (in NLSs) in ${\bf k}$-space. 
Such degeneracies do not originate from high-dimensional irreducible representations 
at such ${\bf k}$ points, but rather 
from the interplay between ${\bf k}$-space topology and symmetry.
Dirac semimetals have been realised in Na$_3$Bi \cite{Liu14s,Xu15} and Cd$_3$As$_2$ \cite{Neupane14,Borisenko14}.
Weyl semimetals have been proposed to exist in pyrochlore iridates \cite{Wan12}, HgCr$_2$Se$_4$ \cite{Xu11}, Te under pressure\cite{Hirayama15}, 
LaBi$_{1-x}$Sb$_x$Te$_3$, LuBi$_{1-x}$Sb$_x$Te$_3$ \cite{Liu14},  
transition-metal dichalcogenides \cite{Sun15,Soluyanov15} 
and SrSi$_2$ \cite{Huang16}.
Consistent with theoretical predictions \cite{Weng15,Huang15}, the 
TaAs class of materials has been experimentally found to 
be Weyl semimetals \cite{Lv15,Xu15b,Lv15b,Xu15c,Yang15}.
}

In the present study, we focus on NLSs.
Two typical origins of the nodal lines are 
(A) mirror symmetry and (B) the $\pi$ Berry phase, as explained in the 
Methods section. 
To our knowledge, proposals for NLSs have thus far been restricted to 
the former mechanism. Dirac NLSs having Kramers degeneracy include
carbon allotropes \cite{Chen15nano,Weng15b}, 
Cu$_3$PdN \cite{Kim15,Yu15}, Ca$_3$P$_2$ \cite{Xie15,Chan15},
LaN \cite{Zeng15}, CaAgX(X=P,As) \cite{Yamakage16} and 
compressed black phosphorus \cite{Zhao15}; similarly, Weyl NLSs, 
which has no Kramers degeneracy, 
 include HgCr$_2$Se$_4$ \cite{Xu11} and TlTaSe$_2$ \cite{Bian15}.
The latter mechanism is also found to occur in some of these materials; 
that is, the nodal line survives despite 
external disruption of mirror symmetry. 
Thus far, no purely topological NLSs 
resulting from the latter mechanism have been proposed.

In the present study we propose on the basis of {\it ab initio} calculation 
that the alkaline-earth metals Ca, Sr and Yb have topological nodal lines
when the spin-orbit  interaction (SOI) 
is neglected. In reality, the SOI is nonzero, especially for Yb, giving rise to 
a small gap along the otherwise gapless nodal lines. 
In fact, the existence of nodal lines
has been observed \cite{vasvari67a,vasvari67b,naumov} and has been used to
explain resistivity data. Nevertheless, its topological origin and its relationship with
surface states remain unexplored. Here, we show their physical origin. 
We also calculate the Zak phase along some reciprocal 
lattice vector and show that the Zak phase is either $\pi$ or 0 depending on the momentum 
regions divided by the nodal lines. 
As the Zak phase is related to polarization, the region with the $\pi$ Zak phase gives rise to  a polarization charge at the surface normal to the  reciprocal 
lattice vector. 
We show that, contrary to common belief, 
the $0$/$\pi$ Zak phase is not related to the absence or presence of surface states.
Unlike insulators, carriers screen this surface polarization charge from the $\pi$ Zak phase, leaving behind surface dipoles.
Finally, we expect the large surface dipoles due to nodal lines to enhance surface
Rashba splitting, possibly contributing to Rashba splitting in Bi/Sr(111) 
(Rashba energy: $E_\text{R}\sim 100$ meV). The 
 large Rashba splitting ($E_\text{R}\sim 200$ meV) on the surface of Bi/Ag(111) 
\cite{ast} is also
attributed to hybridization between the Bi states with 
emergent surface states from the nodal lines in Ag. Thus, the nodal lines are shown to enhance surface Rashba splitting, which is potentially important for spintronics applications.

\begin{figure}[ptb]
\centering 
\includegraphics[clip,width=0.45\textwidth ]{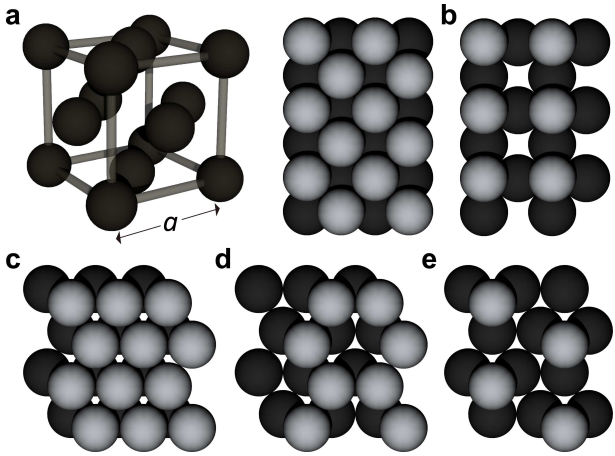} 
\caption{{\bf Bulk and surface structure.}
{\bf a}, Crystal structure of fcc Ca, Sr and Yb, and that of the (001) surface (black circles) with surface atoms (grey circles).
{\bf b}, The same surface orientation but with half of the 
atoms per unit cell on the surface.
{\bf c}, Crystal structure of the (111) surface (black circles) with surface atoms (grey circles).
{\bf d}, {\bf e}, The same surface orientation but with two-thirds and one-third, 
respectively, of the atoms per unit cell on the surface.
}
\label{fcc}
\end{figure} 

\noindent {\bf Results} \\
{\bf Band structures of Ca, Sr and Yb.}
Ca, Sr and Yb are nonmagnetic metals, having a face-centred cubic (fcc) lattice with lattice parameter $a$ (Fig.~\ref{fcc}a). 
The space group of fcc is Fm$\bar{3}$m (No. 225).
At higher pressure, interesting phase transitions have been observed in these metals.
In Yb, the metal-insulator transition occurs at 1.2 GPa~\cite{enderlein13}.
Ca and Sr also exhibit semimetallic behaviour under pressure~\cite{wcwhan69,magnitskaya14}.
The first structural transition from fcc to body-centred cubic (bcc) takes place in 
at 19-20 GPa~\cite{jayaraman63}.
High-temperature superconductivity is observed in Ca at 29 and at 216 GPa after several structural transitions~\cite{sakata11}.

\begin{figure}[ptb]
\centering 
\includegraphics[clip,width=0.45\textwidth ]{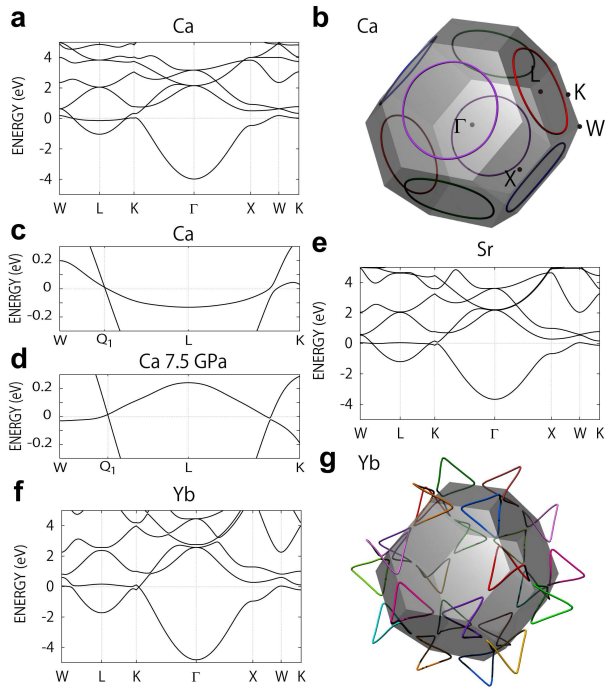} 
\caption{
{\bf Electronic band structure and nodal line.}
{\bf a}, Electronic band structure of Ca in the LDA. 
{\bf b}, Nodal lines and the Brillouin zone of Ca, where identical nodal lines (modulo reciprocal vector) are shown in the same colour. 
{\bf c},{\bf d}, The magnified electronic band structure of Ca at ambient pressure and 
at 7.5 GPa, respectively, in the LDA. 
{\bf e}, {\bf f}, The electronic band structures of Sr and Yb, respectively, 
in the LDA+SO.
{\bf g}, The nodal lines of Yb in the LDA.
The energy is measured from the Fermi level.
}
\label{bulk}
\end{figure} 

We determine their electronic structures by {\it ab initio} calculations, as
explained in the Methods section. 
Figure~\ref{bulk}a shows the electronic structure of Ca
obtained by local density approximation (LDA).
The Brillouin zone is shown as a truncated octahedron in Fig.~\ref{bulk}b.
In a Ca atom, a gap exists 
between the fully occupied $4s$ orbitals and unoccupied $3d$ and $4p$ orbitals.
When the atoms form a crystal, these orbitals form bands with a narrow gap or pseudo-gap near the Fermi level.
The top of the valence band, which is relatively flat near the L points, originates from the $p$ orbital oriented along the [111]-axis having strong $\sigma $ bonding, while most of the 
other valence bands originate from the $s$ and $d$ orbitals.
Around the L points, the relatively flat valence band 
crosses 
the dispersive conduction band. It produces four
nodal lines around the L points within approximately $\pm$ 0.01 eV near the Fermi level, 
as shown in Fig.~\ref{bulk}b. 
(There appear to be eight nodal lines, but the nodal lines in the same 
colour are identical.)
The four nodal lines are mutually related by $C_4$ symmetry, and 
are oriented slightly away from the faces of the first Brillouin zone, except for the points along the L-W
lines (Q$_1$ in Fig.~\ref{bulk}c) because of $C_2$ symmetry.
The nodal lines do not lie on mirror planes; 
therefore, they do not arise from mirror symmetry. 
Because of the topology resulting from the 
$\pi$ Berry phase (see Methods and Supplementary Note 1), 
 closing of the gap is not limited to the 
Q$_1$ points on the L-W lines, instead extending to form nodal lines.  
Indeed, we 
numerically confirmed that the Berry phase around each nodal line is $\pi$.

There are other ways of topological characterisation of  the
nodal lines, distinct from the Berry phase. 
For example, one way of topological characterization 
is the $\mathbb{Z}_2$ indices defined in Ref.~\cite{Kim15}, calculated 
as products of the parity eigenvalues 
of the valence bands at the time-reversal invariant momenta (TRIM). We find that 
all $\mathbb{Z}_2$ indices are even (trivial) for Ca, Sr and Yb.
The existence of the nodal lines is consistent with the trivial $\mathbb{Z}_2$ indices,
since the number of nodal lines between the TRIM is even. 
In that sense, these metals can be called
`weak' NLSs, in analogy with weak topological insulators; namely, the existence of 
nodal lines does not arise from bulk $\mathbb{Z}_2$ indices.  
Another $\mathbb{Z}_2$ index is defined for 
each nodal line in Ref.~\cite{Fang15}. If it is nontrivial, then
it prevents the nodal line from disappearing by itself. In the present case of Ca, this 
$\mathbb{Z}_2$ index is trivial, as shown in detail in the Supplementary Note 4.

At ambient pressure, Ca is not an NLS, because 
the two bands forming the nodal lines both disperse downward 
around the L points (Fig.~\ref{bulk}c). Meanwhile, 
Ca becomes an NLS under pressure, 
as shown in the band structure at 7.5 GPa  in Fig.~\ref{bulk}d;
a similar conclusion has been reached in previous works \cite{vasvari67a,vasvari67b}
without showing the topological origin of the nodal lines.
Here, the pressure increases the energy of the $p$ orbital relative to 
that of the $s$ and $d$ orbitals.

The electronic structure of Sr in the LDA with relativistic effect (LDA+SO, see Methods) is shown in Fig.~\ref{bulk}e.
The SOI is not strong over the entire ${\bf k}$ space.
The band at the L points 
 near the Fermi level is relatively flat compared with that in Ca
because the energy difference between the $5s$, $4d$ and $5p$ orbitals is larger than that between the $4s$, $3d$ and $4p$ orbitals.
With the SOI neglected, the four nodal lines occur around the L points, 
as is the case with Ca.
The nodal line is fully gapped with the SOI;
for example,  the degeneracy on the L-W line splits by approximately 
0.04 eV because of the SOI.

\begin{figure}[ptb]
\centering 
\includegraphics[clip,width=0.45\textwidth ]{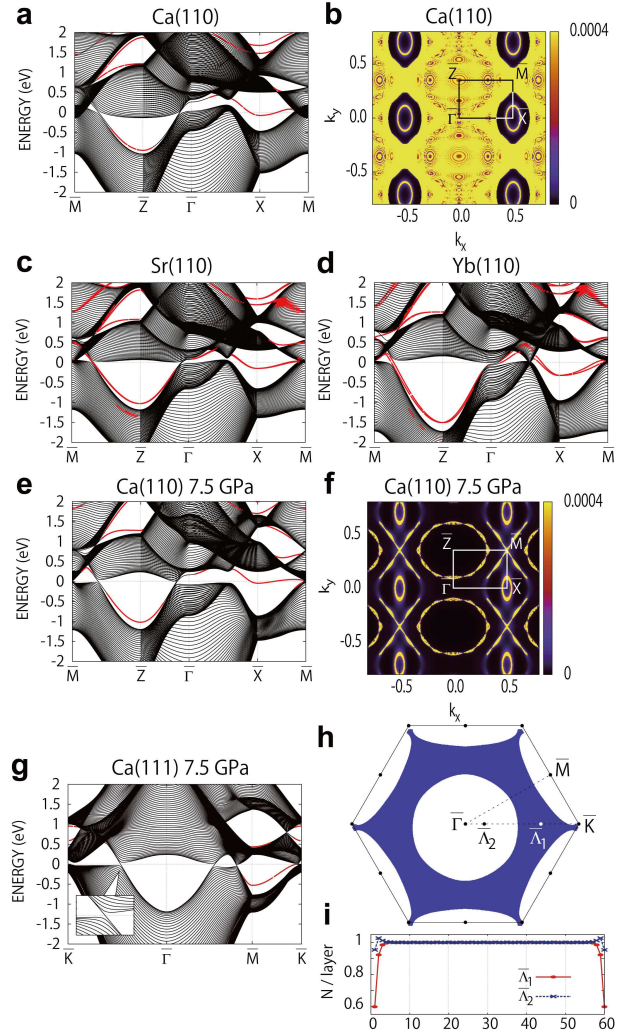} 
\caption{{\bf Topological surface state and surface polarization charge.}
{\bf a}, Electronic band structure for the the(110) surface of Ca  in the LDA.
The symmetry points are $\bar{\rm \Gamma} =$(0, 0, 0), $\bar{\rm X}=(\pi/a)$($\sqrt{2}$, 0, 0), $\bar{\rm Z}=(\pi/a)$(0, 0, 1), and $\bar{\rm M}=(\pi/a)$($\sqrt{2}$, 0, 1).
{\bf b}, Intensity colour plot of the charge distribution in ${\bf k}$ space at the Fermi level
within one atomic layer near the surface
(the unit for the ${\bf k}$ vector is $2\sqrt{2}\pi/a$).
{\bf c}, {\bf d}, Electronic band structures of Sr and Yb for the (110) surface in the LDA+SO, respectively.
{\bf e}, {\bf f}, Results for Ca at 7.5 GPa corresponding to parts  {\bf a} and {\bf b}.
{\bf g}, Electronic band structure for the (111) surface of Ca  in the LDA, with 
states close to the nodal line magnified in the inset.
{\bf h}, Dependence of the Zak phase on the surface momentum ${\bf k}_{\parallel}$. The shaded region represents ${\bf k}_{\parallel}$  with the $\pi$ Zak phase, while other regions represent that with the $0$ Zak phase. 
{\bf i}, Charge profile  for two values of the surface momentum ${\bf k}_{\parallel}=\bar{\Lambda}_1,\bar{\Lambda}_2$
in real space over the thickness direction of the slab. 
The vertical axis represents the charge density per surface unit cell within each atomic layer, measured in units of  electronic charge, $-e$.
In parts {\bf a,c-e,g}, wavefunctions monotonically decreasing toward the bulk region are shown in red.
The energy is measured from the Fermi level.}
\label{surface}
\end{figure} 
Figure~\ref{bulk}f shows the electronic structure of Yb in the LDA+SO.
The energy splitting on the L-W line reaches approximately 0.2 eV.
Unlike Ca and Sr, 
the nodal lines without the SOI in Yb are qualitatively different from those in Ca and Sr;
hybridization between four nodal lines around the X points causes a
Lifshitz transition, i.e. a recombination of the nodal lines, and 12 small nodal lines appear around the W points (Fig.~\ref{bulk}g).
A similar recombination of nodal lines is seen when the lattice constant of the 
Ca crystal is increased in the numerical calculation, which gradually reduces the crystal to the atomic limit. 
The nodal lines form around the W points after recombination, subsequently 
shrinking toward the W points 
and disappearing.
Compared with Ca and Sr, Yb is consequently 
closer to band inversion between the atomic limit and the
crystal.

\noindent
{\bf Zak phase and surface states.} 
We now show the surface states of Ca, Sr and Yb. 
Figure~\ref{surface}a shows the electronic structure of the Ca(110) surface, and
Fig.~\ref{surface}b shows
its charge distribution at the Fermi level.
Similar surface states are also found in the Sr(110) and Yb(110) surfaces
(Fig.~\ref{surface}c, d).
Surface states connecting the gapless points exist near the Fermi level around 
the $\bar{\rm X}$ point, isolated from the bulk states.
In particular,  while the SOI opens a small gap at the nodal lines in Sr and Yb,
the surface states persist by continuity because of nodal lines. 
Out of the four nodal lines in this (110) surface, two 
overlap each other, resulting 
in a projected nodal line around the $\bar{\rm Z}$ point. The other two nodal lines
nearly become segments crossing each other at the $\bar{\rm M}$ point.
Figure~\ref{surface}e is the electronic structure of the Ca(110) surface at 7.5 GPa.
It is in the NLS phase, and the states at the Fermi level 
(Fig.~\ref{surface}f) consist almost exclusively 
of the surface states.

We calculate the Zak phase (the Berry phase),
which 
is an integral of the Berry connection of the bulk wavefunction 
along a certain reciprocal vector ${\bf G}$.
For the calculation, we decompose the wavevector ${\bf k}$ into the components
along ${\bf n}\equiv {\bf G}/|{\bf G}|$ and perpendicular to ${\bf n}$:
${\bf k}=k_{\perp}{\bf n}+{\bf k}_{\|}$, ${\bf k_{\|}}\perp{\bf n}$. 
The integral with respect to $k_{\perp}$ is calculated with fixed ${\bf k}_{\|}$.
The Zak phase is defined in terms of modulo $2\pi$  because of the gauge degree of freedom. 
We focus on the cases without the SOI and 
neglect the spin degeneracy.
As discussed in a previous work~\cite{PhysRevB.48.4442}, 
the Zak phase is related to charge polarization at surface momentum ${\bf k}_{\|}$ for a surface perpendicular 
to ${\bf n}$
(see the Methods section for details).
For example, in a one-dimensional insulating system, the product of the Zak phase and $e/(2\pi)$ is
equal to the polarization, i.e. the amount of surface polarization charge, modulo $e$ \cite{PhysRevB.48.4442}.
Additionally, in a three-dimensional system, if the system at each ${\bf k}_{\|}$ is regarded 
as a one-dimensional system, then the product of 
the Zak phase $\theta({\bf k}_{\|})$ and $e/(2\pi)$
is equal to $\sigma({\bf k}_{\|})$ modulo $e $, where $\sigma({\bf k}_{\|})$ is the amount of surface charge for the one-dimensional system at given ${\bf k}_{\|}$. For an 
insulator, a surface polarization charge density $\sigma_{\rm total}$  at the given surface is
given by $\sigma_{\rm total}=\int\frac{d^2k_{\|}}{(2\pi)^2}\sigma({\bf k}_{\|})$ \cite{PhysRevB.48.4442}.
Because the Berry phase around the nodal line is $\pi$,  
the Zak phase jumps by 
$\pi$ as ${\bf k}_{\|}$ changes across the nodal line.
This is confirmed in our case. The resulting Zak phase is 0 in the entire ${\bf k}_{\parallel}$ space for the (110) surface, because the two nodal lines
out of the four overlap each other while the other two nearly becomes segments. In the (001) surface, the Zak phase is also 0 everywhere, because 
four nodal lines overlap each other in two pairs, and the Zak phase is doubled.
Meanwhile, 
the Zak phase for the (111) surface (Fig.~\ref{surface}{g}) is $\pi$ outside of the nodal line, as shown as the shaded region in Fig.~\ref{surface}.
Within this ${\bf k}_{\parallel}$ region,
 $\sigma({\bf k_{\|}})$ takes a value  $\sigma({\bf k_{\|}}) \equiv e/2\ (\mathrm{mod}\ e)$
\cite{PhysRevB.48.4442},
inevitably leading to a surface polarization charge. 
When the surface termination is
fixed, the value of $\sigma({\bf k_{\|}})$ is determined without the indeterminacy modulo $e$.

For example, the Zak phases for points $\bar{\Lambda}_1$ and $\bar{\Lambda}_2$ in Fig.~\ref{surface}h 
are  $\pi$ and $0$, respectively,  
and the surface polarization charges $\sigma({\bf k}_{\|})$ at these wavevectors are  $e/2$ and 0 (mod $e$). Since  surface states exist
neither at $\bar{\Lambda}_1$ nor at $\bar{\Lambda}_2$, 
this difference in surface polarization charges are due 
to charge distribution of bulk valence bands, 
as demonstrated in Fig.~\ref{surface}i. 
At $\bar{\Lambda}_2$ the charge distribution is almost constant 
even near the surface, whereas at $\bar{\Lambda}_1$ it decreases 
by $\sim (-e)/2$ near each
surface, consistent with the value of $\sigma({\bf k}_{\|})\equiv e/2$ mod $e$.
Since the two surfaces of the slab are equivalent because of inversion symmetry, 
the total charge at $\bar{\Lambda}_1$ is less than that
at $\bar{\Lambda}_2$ by one electron (i.e. charge $(-e)$). This difference is attributed to one state
which traverses the gap from the valence band to the conduction band along 
the $\bar{\rm K}\rightarrow \bar{\Gamma}$ direction (inset of 
Fig.~\ref{surface}g; 
notably, this state is not a surface state, since it disappears at the limit of 
infinite system size. The small gap between $\bar{\rm K}$ and $\bar{\Gamma}$
is a minigap because of the finite-size effect, and this gap goes to zero in the infinite system size.)
To summarize, an $e/2$ surface polarization charge from the $\pi$ Zak phase
is attributed to
bulk states. It holds true 
even when the Zak phase is not quantized; a nonzero Zak phase 
$\theta$ implies that the bulk states have excess polarization charge at the 
surface.

Thus $\sigma({\bf k}_{\|})$ takes the value $\sigma({\bf k}_{\|})=\frac{e}{2}$ in the shaded region in Fig.~\ref{surface}h, 
the area of which is $0.485$ of the total area of the Brillouin zone. Therefore, the 
surface polarization charge density is $\sigma=0.485\cdot \frac{e}{2A_{\mathrm{surface}}}
\sim \frac{0.243e}{A_{\mathrm{surface}}}$ where $A_{\mathrm{surface}}$ is the area of the 
surface unit cell. 
A nonzero 
surface polarization charge  in this centrosymmetric crystal seems unphysical. Nevertheless, 
it does not violate the inversion symmetry, because the amount of 
surface charge is the same for two surfaces of a slab of finite thickness. 
Notably, it is
a surface polarization charge if we regard the system as a collection of one-dimensional systems
for each ${\bf k}_{\|}$. In reality, the excess surface charge are screened by 
free carriers because of the existence of free carriers, as discussed later in this paper.

It is commonly believed that when $\theta({\bf k}_{\parallel})$ equals $\pi $ at some ${\bf k}_{\parallel}$, the drumhead surface states appear at ${\bf k}_{\parallel}$; 
this is indeed the case when the system has chiral symmetry, 
according to a theorem in a previous work~\cite{PhysRevLett.89.077002}.
Here surface states are defined as states having a finite penetration depth in the limit of 
an infinite system size. Nevertheless, it is not always true in general systems. Comparison between the (111) surface states (Fig.~\ref{surface}g) and 
the value of the Zak phase (Fig.~\ref{surface}h) shows no
direct relationship between the absence or presence of the surface states and 
the $0$/$\pi$ Zak phase.
It is in fact natural, as shown by the following discussion. 
Suppose there is a surface state within the gap at ${\bf k}_{\parallel}$ near the nodal line. 
If it is occupied, it then contributes $(-e)$ to the surface polarization charge; if unoccupied, it 
does not contribute. Thus, the presence or absence  or occupancy of 
surface states affects the surface polarization charge 
by an integer multiple of $e$, and 
this cannot account for $e/2$ surface charge from the $\pi$ Zak phase.
Therefore, the $\pi$ Zak phase due to the nodal line
does not imply existence of surface states \cite{Fang15}, and the presence or absence of surface states 
depends on details of the surface being considered.

A previous work has shown that only in systems with chiral symmetry the $\pi$ Zak phase indicates the presence of boundary states (at zero energy) \cite{PhysRevLett.89.077002}. 
This is consistent with our conclusion above. 
When the Zak phase for the bulk occupied bands at certain ${\bf k}_{\parallel}$ value 
is $\pi$, the surface polarization charge for the occupied bands is $\sigma_{\rm occ.}\equiv e/2$ (mod $e$) at this wavevector. 
From the chiral symmetry, the bulk unoccupied bands also have the same surface polarization charge
 $\sigma_{\rm unocc.}=\sigma_{\rm occ.}$. Therefore, the total surface charge
for all of the bulk bands is  $\sigma_{\rm unocc.}+\sigma_{\rm occ.}=2
\sigma_{\rm occ.}\equiv e$ (mod $2e$). Thus, there is an odd number of surface states which accommodate excess electrons at the
surface, and 	there is
a zero-energy surface state due to chiral symmetry. Hence, 
the chiral symmetry is essential to relating the $\pi$ Zak phase to the presence of 
surface states.

\begin{figure}[ptb]
\centering 
\includegraphics[clip,width=0.45\textwidth ]{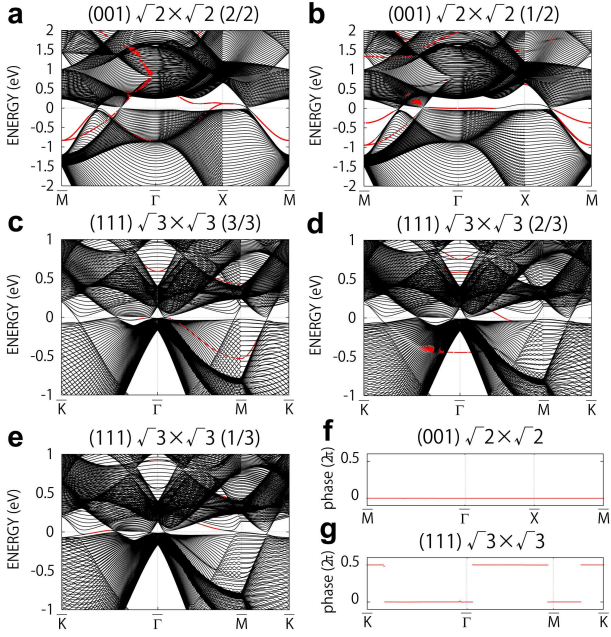} 
\caption{{\bf Topological surface state on long-range superstructure.}
Electronic band structure of Ca at 7.5 GPa, with 
{\bf a} and {\bf b} showing 
 long-range $\sqrt{2}\times \sqrt{2}$ structures on the (001) surface,
and 
{\bf c}-{\bf e} showing 
 long-range $\sqrt{3}\times \sqrt{3}$ structures on the (111) surface
 in the LDA.
The surfaces in {\bf a} and {\bf c} are flat, and those in {\bf b}, {\bf d} and {\bf e} have 
one-half, one-third and two-thirds of the surface atoms per unit cell, respectively
({\bf b}, {\bf d} and {\bf e} correspond to Fig.~\ref{fcc}{\bf b}, {\bf d}, and {\bf e}, respectively).
The symmetry points in {\bf a} and {\bf b} are $\bar{\rm \Gamma} =(2\pi/a)$(0, 0, 0), $\bar{\rm X}=(\pi/a)$(1, 0, 0), and $\bar{\rm M}=(\pi/a)$(1,1, 0); 
those in {\bf c}-{\bf e} are $\bar{\rm \Gamma} =(2\sqrt{2}\pi/\sqrt{3}a)$(0, 0, 0), $\bar{\rm M}=(\sqrt{2}\pi/\sqrt{3}a)$(0, 1, 0), and $\bar{\rm K}=(\sqrt{2}\pi/\sqrt{3}a)$($\sqrt{3}$, 1, 0).
Wave functions decreasing monotonically toward the bulk region are shown in red.
The energy is measured from the Fermi level.
{\bf f}, The Zak phase from the occupied bands for both {\bf a} and {\bf b}. 
{\bf g}, The Zak phase from the occupied bands for {\bf c}-{\bf e}. 
}
\label{long_periodic}
\end{figure} 

We emphasize that this surface charge is determined from the bulk bands. This can be
seen by considering the surfaces with periodic depletion of some atoms, such that the surface forms a superstructure. For the (001) surface, we consider two patterns for the $\sqrt{2}\times\sqrt{2}$ 
superstructure, as shown in Figs.~\ref{fcc}b and c. Figure~\ref{fcc}b represents the perfect surface; in Fig.~\ref{fcc}c, half of the surface atoms per unit cell are present. 
For the (111) surface, we consider three patterns for the$\sqrt{3}\times\sqrt{3}$ 
superstructure, as shown in Figs.~\ref{fcc}c-e. Fig ~\ref{fcc}c represents the perfect surface; 
in Figs.~\ref{fcc}d and e, two-thirds and one-third, respectively, of the surface atoms per unit cell are present.
The band structures are shown in Figs.~\ref{long_periodic}a-e. 
We also found that for the (001) surface, two patterns for the $\sqrt{2}\times\sqrt{2}$ (a and b) give the same  Zak phase (see Fig.~\ref{long_periodic}f). 
Similarly, three patterns for the $\sqrt{3}\times\sqrt{3}$ 
superstructure for the (111) surface (c-e) give the same Zak phase (see
 Fig.~\ref{long_periodic}g).
Although the Zak phase generally
depends on the surface termination via the choice of the unit cell  \cite{PhysRevB.88.245126},
it is unchanged in the present case when the surface termination is changed for a fixed surface orientation,
as shown 
in the Supplementary Note 3. 
This result is 
natural because the Zak phase and resulting surface polarization charge (modulo $e$) are bulk quantities, 
and are independent of formation of surface superstructure.

\noindent{\bf Nodal lines and Rashba splitting.} 
Thus far we have shown that the nodal lines affect the surface in 
two ways. One is through the surface charge via the Zak phase.
 Across the nodal lines the 
Zak phase changes by $\pi$, and regions with $\pi$ Zak
phase (having $e/2$ (mod $e$) surface charge) encircled by the nodal lines appear, giving rise to a 
$\pm e/2$ surface polarization charge. The other route is through the 
surface states. In some NLSs, the drumhead surface state 
emerges. These two induce an appreciable 
dipole at the surface and may enhance surface Rashba splitting, 
as we show below.

This property is unique to materials with nodal lines. For
a comparison, let us first consider an insulator with inversion and time-reversal symmetries. Consequently, 
the Zak phase 
satisfies $\theta({\bf k}_{\|})=0\ {\rm or}\ \pi$ (mod $2\pi$) and it gives $\sigma({\bf k}_{\|})=Ne/2$ (where $N$ is an integer; see Supplementary Note 2). Because 
$\theta({\bf k}_{\|})$ is continuous for all ${\bf k}_{\|}$, $N$ is
common for all  ${\bf k}_{\|}$. 
Thus far, no insulator with $N\neq 0$ is known, to our knowledge, possibly 
because of its instability due to the huge polarization charge at the surface. A
nonzero even value of $N$ is not topologically protected, and it is easily reduced to $N=0$. Meanwhile, an odd value of $N$ means that dangling bonds, which are expected to be  unstable, cover the surface. Thus, only the case  $N=0$ is physically realizable.

In materials with nodal lines, the Zak phase jumps by $\pi$ at the nodal lines;
therefore, there are always two types of regions, one with $\theta({\bf k}_{\|})\equiv 0$ (mod $2\pi$) and
one with $\theta({\bf k}_{\|})\equiv \pi$ (mod $2\pi$). The latter region inevitably leads to an appreciable polarization, as is exemplified by the Ca surface. In NLSs, 
the bulk charges eventually screen the polarization, but large deformation of the 
lattice structure and electronic relaxation (i.e. screening) occur.
In the present case, carriers screen 
the surface polarizations, leaving behind dipoles at
the surface. As roughly estimated for calcium (see Supplementary Note 5), 
the dipole density per surface unit cell is $\sim 5\times 10^{-21}\mathrm{C}\cdot \mathrm{nm}$, the potential dip is $\sim -0.8$eV at the surface, and the electric field at the surface is $\sim 6.4\mathrm{V}\ \mathrm{nm}^{-1}$.

\begin{figure}[ptb]
\centering 
\includegraphics[clip,width=0.45\textwidth ]{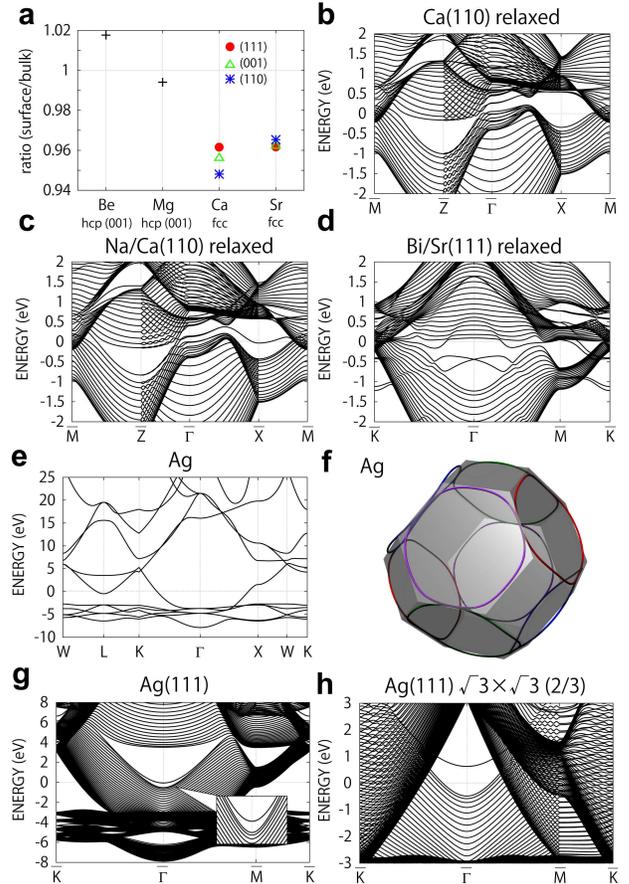} 
\caption{{\bf Nodal line and Rashba effect.}
{\bf a}, Ratio of the distance between the layers at the surface to that of bulk (20 layers).
{\bf b}, {\bf c}, Electronic band structure of Ca for the (110) surface without 
Na and covered with Na, respectively, in the LDA.
{\bf d}, Electronic band structure of Sr for the (111) surface covered with Bi in the LDA+SO.  
Structures in {\bf a}-{\bf d} are optimized by GGA .
{\bf e}, Electronic band structure of Ag.
{\bf f}, Nodal lines around 5 eV and the Brillouin zone of Ag, where the nonequivalent nodal line is shown in the different color. 
{\bf g}, {\bf h}, Electronic band structure of Ag, without long-range $\sqrt{3}\times \sqrt{3}$ structure and with long-range $\sqrt{3}\times \sqrt{3}$ structure, respectively,
on the (111) surface.
The energy is measured from the Fermi level.
}
\label{polar}
\end{figure}

Figure~\ref{polar}a shows the ratio of the interlayer distance at the surface to that of the bulk for the several surfaces of fcc Ca and Sr and the (001) surface of hexagonal close-packed (hcp) Be and Mg.
The Ca and Sr surfaces, having nodal lines,
are compressed near the surface by around 4 \% (equivalent to that in the bulk of Ca at $\sim 2$ GPa).
This large compression in Ca and Sr may be associated with the surface charge 
induced by the nodal lines. 
From the above argument, 
the effect of the large charge imbalance at the surface is prominent only when the nodal lines are almost at the Fermi energy and no other Fermi surfaces exist. 
As shown in the Supplementary Note 5, carriers in the semimetal screen the surface charge through a screening length on the order of nanometers. Thus, 
 the screening in this case is poor because the NLS has a small number of 
carriers, leaving behind an appreciable dipole moment at the surface after the 
screening. 
On the other hand, if there are carriers other than those forming the nodal line, 
then the screening effect is much more prominent, and dipoles at the surface are small. 
In Be and Mg, there are nodal lines \cite{Mikitik,Li16}
away from the Fermi level (at 0.0--1.1 eV in Be and at 0.6--1.1 eV in Mg),
and the density of states at the Fermi energy is large, because of the large Fermi surface; 
therefore, the compression of the lattice is small.
In addition to the (111) surface of Ca, 
the (001) and (110) surfaces also show large deformations, because of 
large surface charges,
where some ${\bf k}$ points with 0 $\equiv 2\pi\ (\mathrm{mod}\ 2\pi)$ Zak phase are shown to have $\sigma({\bf k}_{\|})=-e$ surface charge in our calculation (not shown).
Thus the large compression on the Ca surface is almost independent of the surface termination.

We find that the electronic relaxation alters
the surface state dispersions  (Fig.~\ref{polar}b) relative to the case with no
relaxation (Fig.~\ref{surface}a). Meanwhile, 
the energies of some surface states are lowered by 
$\sim 1$eV. This effect is attributed to the potential dip at the surface due to 
surface charge, estimated as $\sim -0.8$eV.
This interpretation is also confirmed by subsequent calculation. 
When the instability originating from the surface charge
is eliminated by covering the surface with alkali metal with low electronegativity, the surface states (Fig.~\ref{polar}c)
emerge as almost the same as 
the original surface without relaxation (Fig.~\ref{surface}a).

We expect that such huge polarization charge due to the nodal lines would enhance
 Rashba spin-splitting at the surface.
In Fig.~\ref{polar}d, we show the Rashba splitting in a Bi monolayer on the
Sr(111) surface.
The Rashba splitting of the 
surface Bi $6p$ bands near the Fermi level is as large as $E_\text{R}\sim 100$ meV. 
This large Rashba splitting may be partially attributed to the nodal lines;
the potential dip at the surface due to the nodal lines gives rise to an  
additional strong electric field within the Bi layer, roughly on the order of $1{\rm V}/5{\rm \AA}\sim 2$V nm$^{-1}$. 
Such a strong additional electric field is expected to enhance the
Rashba splitting. In addition to the nodal lines, 
the difference in the electronegativity between Sr and Bi may also enhance 
the Rashba splitting. 
Nonetheless, evaluating the contribution of the nodal line to the Rashba
splitting is difficult,
because the magnitude of the Rashba splitting is determined by the electric field very close to surface nuclei, and
by the asymmetry of the wavefunctions of the surface atoms \cite{Bihlmayer}. 
For comparison we give some examples: The Rashba parameter increases by 0.005 and 0.011 nm$\cdot$eV under an
external electric field $E=$4.0 V nm$^{-1}$ on the Au(111) \cite{Gong} and KTaO$_3$(001)
surfaces \cite{Shanavas}, respectively. 
On the Bi/Sr(111) surface, the Rashba parameter
for the Bi 6$p_z$ band is 0.071 nm$\cdot$eV, a part of which is attributed 
to the nodal lines.

For comparison, we discuss the Bi/Ag(111) surface, which is known to 
show large Rashba splitting \cite{ast}.
We can also attribute this to nodal lines in Ag, but to a mechanism different from 
that in Bi/Sr(111). 
The conduction band structure of fcc Ag in Fig~\ref{polar}e 
closely resembles that of Ca, Sr and Yb; Ag also has topological Dirac nodal lines around 5 eV (Fig.~\ref{polar}f).
The Dirac nodal lines give rise to surface states for the (111) surface, 
which are visible around $\bar{\Gamma}$ near the Fermi energy ,as shown 
in Fig.~\ref{polar}g. 
The well-known large Rashba splitting is realized when
Bi atoms replace one-third of the surface Ag atoms, forming the 
$\sqrt{3}\times\sqrt{3}$ structure \cite{ast}. 
To establish the origin of the large Rashba splitting, in Fig.~\ref{polar}h we show the band structure for
the $\sqrt{3}\times\sqrt{3}$ Ag(111) surface
with one-third of the surface Ag atoms depleted (Fig.~\ref{fcc}d);
surface states also exist around 0.6 eV, which is higher than those in Fig.~\ref{polar}g 
because of depletion of some of the bonds. 
The Bi/Ag(111) surface is realized by adding Bi atoms to this surface,
and 
its surface states around the Fermi energy are
formed by covalent bonding between Bi atoms and Ag surface states~\cite{bian2013}.
Therefore, the Ag surface states around 0.6 eV in Fig.~\ref{polar}h
become stable by hybridization with the Bi $sp_z$ band and thus exhibit the large Rashba splitting. 

Therefore, the nodal lines near the Fermi level enhance the Rashba splitting at the 
surface in two ways: one enhances via the surface charge, arising from the $\pi$ 
Zak phase, and the other is via hybridization with 
the emergent surface states from the nodal lines. The first scenario applies to 
Bi/Sr(111), whereas the second scenario occurs in 
Bi/Ag(111).

\noindent
{\bf Discussion}\\
To summarize, 
{\it ab initio} calculation shows that the alkaline-earth metals Ca, Sr and Yb have topological nodal lines near the Fermi level in the 
absence of the SOI. 
Ca becomes an NLS at high pressure.
The nodal lines do not 
lie on the mirror planes and have a solely topological origin characterized by
the $\pi$ Berry phase. Consequently, the calculated Zak phase is either 0 or $\pi$, depending on the momentum. The $\pi$ Zak phase leads to a 
polarization charge at its surface region. After screening by carriers, the surface charge induces 
surface dipoles and a surface potential dip. The SOI gives rise to 
a small gap at the nodal line, while
the surface states survive. Surface termination affects the surface states, while the Zak phase remains unaffected.
In materials with nodal lines, both the large surface polarization charge and the
emergent drumhead surface states can enhance Rashba splitting on the surface, 
as demonstrated in Bi/Sr(111) and in Bi/Ag(111).

{\small 
\noindent
{\bf Methods}\\
\noindent
{\bf Details of the first-principles calculation.} 
We
calculate the band structures within the density functional theory.
We use  the \textit{ab initio} code QMAS (Quantum MAterials Simulator)  
and OpenMX. 
The electronic structure is calculated using
LDA with and without the relativistic effect (LDA and LDA+SOI, respectively).
We also optimize the lattice parameter under pressure based on the generalized gradient approximation (GGA).The lattice parameter $a$ for Ca/Sr/Yb/Ag is 5.5884/6.0849/5.4847/4.0853 \AA, respectively.
The plane-wave energy cut-off is set to 40 Ry for Ca, Sr and Ag and 50 Ry for Yb.
The $12\times 12\times 12$ regular ${\bf k}$-mesh including the $\Gamma$ point, with the Gaussian broadening of 0.025 eV, is employed. 
We construct an $spd$ model for Ca, Sr, Yb and Ag from the Kohn-Sham bands, using the maximally localized Wannier function~\cite{marzari97,souza01}.
Since the $4f$ orbital in Yb is fully occupied~\cite{matsunami08}, we first construct
 the $spdf$ model for Yb and disentangle out the $4f$ orbital from the model.
We take 90 atoms  on the (110) surface, and 60 atoms on the (111) and (001) surfaces
in the calculation of the electronic band structure  (see Fig.~\ref{surface}b,d).
Similarly, we use 
120 atoms on the (001) surface with long-range $\sqrt{2}\times \sqrt{2}$ structure,
and 180 atoms on the (111) surface with long-range $\sqrt{3}\times \sqrt{3}$ structure
 (see Fig.~\ref{long_periodic}c,e,f). 
We also use 
30 atoms on the (110) surface with lattice relaxation, and 
20 atoms on the (111) surface with lattice relaxation
 (see Figs.~\ref{polar}b,c,d).
Lastly, we use
45 atoms on the (111) surface without 
long-range $\sqrt{3}\times \sqrt{3}$ structure of Ag and
135 atoms with long-range $\sqrt{3}\times \sqrt{3}$ structure of Ag (see Fig.~\ref{polar}g,h).
The electronic structure of Ag is calculated in the LDA (Fig.~\ref{polar}e, g and h).
We take the vacuum region with thickness 20 $\AA$ for the slab calculation.
Density of states at the Fermi level on the (110) surface is calculated by the surface Green's function~\cite{turek97,dai08}
in which the system contains 720 atoms.
(The electronic structure and lattice constant in Fig.~\ref{polar} are calculated directly from the Kohn-Sham Hamiltonian, and other results in Figs.~\ref{bulk}-\ref{polar} are obtained via the Wannier function.)

\noindent
{\bf Conditions for emergence of nodal lines.} 
Mechanisms of the emergence of nodal lines involve either (A) mirror symmetry or (B) the $\pi$ Berry phase. For systems with mirror symmetry (A),
the states on the mirror plane can be classified into two classes based on 
the mirror eigenvalues.
If the valence and conduction bands have different mirror eigenvalues, then the two bands
have no hybridization on the mirror plane, even if the two bands approach and cross. 
This results in a degeneracy along a loop on the mirror plane. This line-node degeneracy is 
protected by mirror symmetry; once the mirror symmetry is broken, the degeneracy is
lifted in general. 
The second mechanism (B) occurs in spinless systems with inversion and time-reversal 
symmetries. The Berry phase
around the nodal line is $\pi$. 
In spinless systems with inversion and time-reversal symmetries, the Berry
phase along any closed loop is quantized as an integer multiple of $\pi$; therefore, 
the nodal line is topologically protected. Hence, closing of the gap 
occurs not at an isolated point in ${\bf k}$ space, 
but along a curve (i.e. a nodal line) in general. 
If the band gap closes at some ${\bf k}$ by symmetry (such as twofold rotation), 
the closing of the gap is not limited to a single value of the wavevector ${\bf k}$;
instead, it extends 
along a nodal line in ${\bf k}$ space, as 
explained in detail in the Supplementary Note 1. 
 In some materials the two mechanisms coexist, whereas in others 
the nodal line originates from only one of these mechanisms.

\noindent
{\bf Calculation of the Zak phase along a reciprocal vector ${\bf G}$.} 
We separate the Bloch wavevector 
${\bf k}$ into the components
along ${\bf n}\equiv {\bf G}/|{\bf G}|$ and perpendicular to ${\bf n}$:
${\bf k}=k_{\perp}{\bf n}+{\bf k}_{\|}$, ${\bf k_{\|}}\perp{\bf n}$.
The integral with respect to $k_{\perp}$ is calculated with fixed ${\bf k}_{\|}$.
For each value of  ${\bf k}_{\parallel} $,  one can define
the Zak phase by
\begin{align}
\theta ({\bf k}_\parallel )=-i\sum_{n}^{{\rm occ.}}\int _{0 }^{2\pi/a_{\perp} } d k_{\perp} 
\left\langle{u_n({\bf k})}\right| \nabla_{k_{\perp}} \left|{u_n({\bf k})}\right\rangle,
\label{eq:Zak}\end{align}
where 
 $u_n({\bf k})$ is a bulk eigenstate in the $n$-th band, the sum is over the occupied states, and $a_{\perp}$ is the 
size of the unit cell along the vector ${\bf n}$ (see Supplementary Note 2). The gauge is taken to be 
$u_n({\bf k})=u_n({\bf k}+{\bf G})e^{i{\bf G}\cdot{\bf r}}$. The Zak phase is defined in terms of modulo 
$2\pi$.
Under both inversion and time-reversal symmetries,
the Zak phase is shown to take a quantized value 0 or $\pi \pmod{2\pi}$ \cite{Chan15,PhysRevB.88.245126}, as is also shown in the Supplementary Note 2.
All of the cases presented in this paper satisfy the symmetry conditions for 
the Zak phase to be  quantized as 0 or $\pi$.
In one-dimensional insulators, the product of Eq.~(\ref{eq:Zak}) and $e/(2\pi)$  
is equal to the polarization, i.e. the amount of the surface charge modulo $e$ \cite{PhysRevB.48.4442}.
In three-dimensional insulators, the surface polarization charge $\sigma({\bf k}_{\|})$ at ${\bf k}_{\|}$
is calculated by multiplying the Zak phase by $e/(2\pi)$ 
and by integrating over the momentum ${\bf k}_{\|}$ perpendicular to ${\bf G}$. 
Thus this quantity is related to the surface polarization charge for the surface perpendicular 
to ${\bf n}$ \cite{PhysRevB.48.4442}.




\vspace{2mm}

\noindent
{\bf Acknowledgement}\\
We thank Shoji Ishibashi for providing us with the {\it ab initio} code (QMAS) and pseudopotentials.
This work was supported by the Grant-in-Aid for Scientific Research
(nos. 26287062 and 26600012), 
by the Computational Materials Science Initiative (CMSI), 
Japan,
and by the MEXT Elements Strategy Initiative to Form Core Research Center (TIES).

\vspace{2mm}

\noindent
{\bf Additional information}\\
The authors declare no competing financial interests.

\vspace{2mm}

\noindent
{\bf Author contribution}\\
All authors contributed to the main contents of this work.
M.H. performed  
the {\it ab initio} calculation with contributions from T.M.
R.O. constructed the arguments on the Zak phase and surface states. 
S.M. conceived and supervised the project.  
M.H., R.O. and S.M. drafted the manuscript.
T.M. provided critical revisions of the manuscript. 

\clearpage
\noindent
{\Large Supplementary Information}

\renewcommand{\theequation}{S.\arabic{equation}}
\setcounter{equation}{0}
\renewcommand{\tablename}{Table S}
\setcounter{table}{0}
\renewcommand{\figurename}{FIG. S}
\setcounter{figure}{0}

\noindent
{\bf Supplementary Note 1. Nodal lines stemming from the $\pi$ Berry phase}

As explained in the Methods section, nodal lines
in spinless systems with inversion and time-reversal 
symmetries can originate from the $\pi$ Berry phase. 
Here, we explain this mechanism.
The Berry phase $\phi (\ell)$ along a loop $\ell$ 
is defined as
\begin{align}
\phi (\ell)=-i\sum_{n}^{\text{occ.}}\int_{\ell} d {\bf k}\cdot 
\left\langle{u_n({\bf k})}\right| \nabla_{\bf k} \left|{u_n({\bf k})}\right\rangle,
\label{eq:phi}
\end{align}
where
 $u_n({\bf k})$ is a bulk eigenstate in the $n$-th band, and 
the sum is over the occupied states.
We define the Berry phase in terms of modulo $2\pi$ because it can change by an integer multiple of 
$2\pi$ under gauge transformation. 
Under a product of time-reversal and spatial inversion operations, 
this quantity can be transformed into $-\phi(\ell)$:
\begin{equation}
\phi(\ell)\equiv -\phi(\ell) \ \ ({\rm mod}\ 2\pi).
\end{equation}
This leads to quantization of the Berry phase $\phi$ as $\phi(\ell)\equiv 0\ {\rm or}\ \pi\  ({\rm mod}\ 2\pi)$. 
Under a continuous change of $\ell$ in the ${\bf k}$ space, 
a jump of $\phi(\ell)$ occurs only when the band gap closes.
Therefore, if the Berry phase $\phi(\ell)$ is $\pi$ (mod $2\pi$) for a certain value of 
$\ell$, then the loop $\ell$ cannot continuously 
deform to a point without closing a gap. 
This condition means that closing of the gap occurs along a loop (nodal line) 
in ${\bf k}$ space, and that this loop is linked to $\ell$.

We also describe in terms of an effective model a mechanism for the appearance
of the nodal line. 
An effective model for a single valence band and a single conduction band is 
generally described as follows:
\begin{align}
&H({\bf k})=\left(
\begin{array}{cc}
a_0({\bf k})+a_z({\bf k}) &a_x({\bf k})-i a_y({\bf k})\\
a_x({\bf k})+i a_y({\bf k})&
a_0({\bf k})-a_z({\bf k}) 
\end{array}
\right)\nonumber \\
&\ \ \ =a_0+a_x\sigma_x+a_y\sigma_y+a_z\sigma_z.
\end{align}
We assume that $a_i({\bf k})$ ($i=0,x,y,z$) is a continuous function of 
${\bf k}$. 
In the presence of both inversion and time-reversal symmetries, 
we obtain $a_y=0$ (after an appropriate unitary transformation in some 
cases).
Thus we have $H({\bf k})=a_0+a_x\sigma_x+a_z\sigma_z$. The 
band gap closes
only if the following conditions are satisfied simultaneously:
\begin{align}
a_x({\bf k})&=0, \label{eq:ax} \\
a_z({\bf k})&=0. \label{eq:az}
\end{align}
Each of these equations determines 
a surface in ${\bf k}$ space, and their intersection gives a nodal line 
in ${\bf k}$ space. 
In this case, a Berry phase $\phi (\ell)$ (Supplementary Eq.~(\ref{eq:phi})) 
along a loop $\ell$ around the nodal line,
 is calculated to be equal to 
a change of phase of ${\rm arg}\frac{a_z}{a_x}$. This is found to be $\pi$ (mod $2\pi$), 
in agreement with the discussion in the previous paragraph.

\noindent
{\bf Supplementary Note 2. Zak phase and polarization}

\noindent
{\bf 2.1 Decomposition of the wavevector components with respect to the surface Brillouin zone}

In preparation for the calculation in the next subsection, we show the 
formula for the decomposition of the wavevector ${\bf k}$ into the surface normal $k_{\perp}$ and 
the directions along the surface ${\bf k}_{\|}$. 
For the calculation of the Zak phase 
we use the formula 
\begin{align}
\theta ({\bf k}_\parallel )=-i\sum_{n}^{\text{occ.}}\int_0^{b_{\perp}}d k_{\perp} 
\left\langle{u_n({\bf k})}\right| \nabla_{k_{\perp}} \left|{u_n({\bf k})}\right\rangle,
\label{eq:theta}
\end{align}
where $b_{\perp}$ is the width of the Brillouin zone along the direction perpendicular to the surface. 
However, defining the integration region when the primitive vectors are not orthogonal to each other is not straightforward.
As we discussed in the main text, we consider the superstructure of the surface; 
furthermore,
the 
primitive vectors may differ from the standard choice. Below, we formulate the Brillouin
zone of the crystal, which takes into account the surface periodicity. 

Let ${\bf a}_{1\|}$ and ${\bf a}_{2\|}$ denote the primitive vectors along the surface. If a surface superstructure is
formed, then these primitive vectors should be chosen to comply with the superstructure. We then introduce another vector ${\bf a}'$ such that 
$\{{\bf a}_{1\|},{\bf a}_{2\|},{\bf a'} \}$ is a set of three-dimensional (3D) primitive vectors  which takes into account 
the surface superstructure. 
Thus, it is not necessarily the primitive vectors of the 3D bulk crystal, but it is 
the minimal set of translation vectors which respects surface superstructure.

We then  take the primitive reciprocal vectors $\{{\bf b}_{1},{\bf b}_{2},{\bf b}_{\perp} \}$ 
from 
$\{{\bf a}_{1\|},{\bf a}_{2\|},{\bf a}' \}$:
\begin{align}
&{\bf b}_{1}
=
2\pi
\frac{{\bf a}_{2\|}\times{\bf a'} }{({\bf a}_{1\|}\times {\bf a}_{2\|})\cdot{\bf a'} }, \\
&{\bf b}_{2}
=
2\pi
\frac{{\bf a'}\times {\bf a}_{1\|} }{({\bf a}_{1\|}\times {\bf a}_{2\|})\cdot{\bf a'} }, \\
&{\bf b}_{\perp}
=
2\pi
\frac{{\bf a}_{1\|}\times {\bf a}_{2\|}}{({\bf a}_{1\|}\times {\bf a}_{2\|})\cdot{\bf a'} }
\end{align}
We note that ${\bf b}_{\perp}$ is normal to the surface, whereas ${\bf b}_{1}$ and 
${\bf b}_{2}$ are not necessarily along the surface (see Supplementary Figure~\ref{translation}). 
We then project ${\bf b}_{1}$ and 
${\bf b}_{2}$ onto the surface:
\begin{align}
&{\bf b}_{1\|}={\bf b}_{1}-\frac{{\bf b}_{1}\cdot{\bf n}}{{\bf n}\cdot{\bf n}}{\bf n},\\
&{\bf b}_{2\|}={\bf b}_{2}-\frac{{\bf b}_{2}\cdot{\bf n}}{{\bf n}\cdot{\bf n}}{\bf n},
\end{align}
where ${\bf n}$ is the unit vector normal to the surface.
It then follows that 
\begin{equation}
{\bf a}_{i\|}\cdot{\bf b}_{j\|}=2\pi\delta_{ij}\ \ \ (i,j=1,2);
\end{equation}
therefore the set $\{{\bf b}_{1\|},{\bf b}_{2\|}\}$ is a set of two-dimensional (2D) primitive
reciprocal vectors for the surface, corresponding to the 2D primitive vectors along the surface, 
$\{{\bf a}_{1\|},{\bf a}_{2\|} \}$. 
Furthermore, the 3D Brillouin zone, which is a parallelogram spanned by $\{{\bf b}_{1},{\bf b}_{2},{\bf b}_{\perp} \}$, is 
equivalent to the parallelogram spanned by 
$\{{\bf b}_{1\|},{\bf b}_{2\|},{\bf b}_{\perp} \}$, with 
${\bf b}_{\perp}$ perpendicular to the surface.
Therefore,  we take
$k_{\perp}$ from zero to $b_{\perp}$
in Supplementary Eq.~(\ref{eq:theta}), while ${\bf k}_{\|}$ takes a 2D wavevector 
within the 2D Brillouin zone spanned by $\{{\bf b}_{1\|},{\bf b}_{2\|}\}$.
We note that $b_{\perp}$ is equal to $2\pi/a'_{\perp}$, where $a'_{\perp}$ is a
surface-normal component of ${\bf a}'$.
\begin{figure}[ptb]
\centering 
\includegraphics[clip,width=0.45\textwidth ]{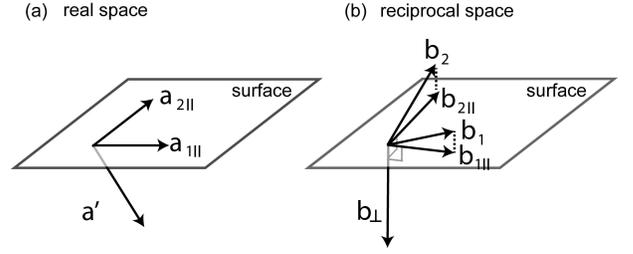} 
\caption{{\bf Primitive vectors.}
Primitive vectors in {\bf a}, {\bf b}, real space and reciprocal space, respectively, 
used in our calculation}
\label{translation}
\end{figure} 

\noindent
{\bf 2.2 Symmetry properties of the Zak phase in three dimensions}

Here we note on symmetry properties of the Zak phase.
We first review the results shown in previous works \cite{Zak,PhysRevB.48.4442},
and then we discuss results for nodal-line semimetals. 
We first rewrite the Schr\"{o}dinger equation ${\cal H}\psi_{{\bf k}}=E_{{\bf k}}\psi_{{\bf k}}$ in terms of the
Bloch wavefunction, where ${\cal H}$ is the Hamiltonian, ${\bf k}$ is the Bloch wavevector, 
$\psi_{{\bf k}}$ is the wavefunction, and $E_{{\bf k}}$ is the energy.
Throughout the paper, we adopt the gauge 
\begin{equation}
\psi_{{\bf k}+{\bf G}}=\psi_{{\bf k}},
\label{eq:gauge}
\end{equation}
where ${\bf G}$ is any reciprocal lattice vector \cite{Zak,PhysRevB.48.4442}; this
choice of gauge is necessary for relating the Zak phase $\theta({\bf k}_{\|})$ to 
polarization.
We then obtain
\begin{equation}
\hat{H}_{{\bf k}}u_{\bf k}=E_{{\bf k}}u_{\bf k},
\end{equation}
where $\psi_{{\bf k}}=u_{{\bf k}}e^{i{\bf k}\cdot{\bf r}}$ and 
$\hat{H}_{{\bf k}}\equiv e^{-i{\bf k}\cdot{\bf r}}{\cal H} e^{i{\bf k}\cdot{\bf r}}$.
The choice of gauge in Supplementary Eq.~(\ref{eq:gauge}) is rewritten as 
$u_{{\bf k}}=u_{{\bf k}+{\bf G}}e^{i{\bf k}\cdot{\bf r}}$.


When the inversion and time-reversal symmetries are present in spinless systems, 
the Zak phase $\theta(C)$ around any closed loop $C$ is quantized as follows:
\begin{align}
\theta(C)\equiv -i\sum_{n}^{\text{occ.}}\oint_C d {\bf k}\cdot
\left\langle{u_n({\bf k})}\right| \nabla_{{\bf k}} \left|{u_n({\bf k})}\right\rangle=
n\pi\ \ (n:\ \mathrm{integer}).
\end{align}
In particular, the Berry phase around the nodal line (contour c in Supplementary Figure~\ref{Zak phase}) is $\pi$; because of the
above quantization, the nodal line is topologically protected.
Thus the Zak phases (Berry phase) along the surface normal $k_{\perp}$ changes by $\pi$, when ${\bf k}_{\|}$ is changed across the projection
of the nodal line onto the surface. In Supplementary Figure~\ref{Zak phase} 
the Zak phases for the paths a and b differ by $\pi$, which is the Berry  
phase around the loop c around the nodal line.
\begin{figure}[ptb]
\centering 
\includegraphics[clip,width=0.45\textwidth ]{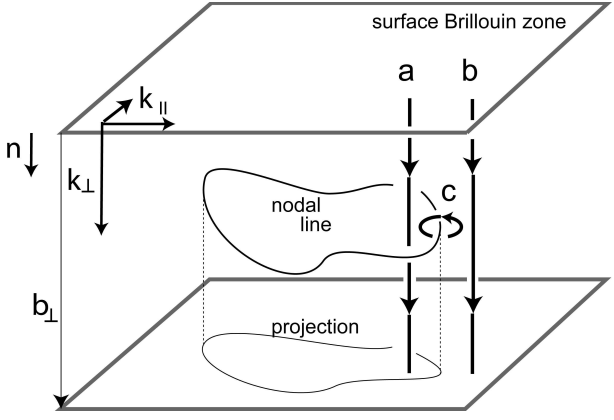} 
\caption{{\bf Zak and Berry phases.}
Relation between the Zak phases (Berry phase) along the surface normal 
(paths a, b) and the
Berry phase around the nodal line. }
\label{Zak phase}
\end{figure} 


In the following we show the effects of the symmetries of the system
 on $\theta ({\bf k}_{\|})$ 

\vspace{1ex}

\noindent
\underline{(i)\ Inversion symmetry}

When the system has inversion symmetry, we have
\begin{equation}
[{\cal P},{\cal H}]=0 \ \Rightarrow \ {\cal P}\hat{H}_{{\bf k}}{\cal P}^{-1}=\hat{H}_{-{\bf k}},
\end{equation}
where ${\cal P}$ is the inversion operator.
We can then derive the 
relationship for the Zak phase:
\begin{equation}
\theta({\bf k}_{\|})\equiv -\theta(-{\bf k}_{\|})\ (\rm{mod}\ 2\pi).
\label{eq:inv}
\end{equation}
Compared with the present result, the results in a previous work 
\cite{PhysRevB.88.245126} contains 
an additional term. This is due to the choices of gauge; the above study~\cite{PhysRevB.88.245126} adopted the gauge $u_{{\bf k}}=u_{{\bf k}+{\bf G}}$, 
whereas we adopt the gauge in Supplementary Eq.~(\ref{eq:gauge}), which is directly related to
the polarization. 

\vspace{1ex}

\noindent
\underline{(ii) Time-reversal symmetry}

When the system has time-reversal symmetry, we have
\begin{equation}
[{\cal K},{\cal H}]=0 \ \Rightarrow \ {\cal K}\hat{H}_{{\bf k}} {\cal K}=\hat{H}_{-{\bf k}},
\end{equation}
where ${\cal K}$ is the complex conjugation.
Here we focus on spinless systems, for which the time-reversal operation is 
represented as ${\cal K}$. The Zak phase then satisfies 
\begin{equation}
\theta(-{\bf k}_{\|})\equiv \theta({\bf k}_{\|})\ (\rm{mod}\ 2\pi).
\label{eq:TR}
\end{equation}
This is the same as that in the above study~\cite{PhysRevB.88.245126}, although the
gauges are different from ours.

\vspace{1ex}

\noindent
\underline{(iii) Inversion and time-reversal symmetries}

When the system has both time-reversal and inversion symmetries, 
Eqs.~(\ref{eq:inv}) and (\ref{eq:TR}) yield from the results in (i) and (ii)
\begin{equation}
\theta({\bf k}_{\|})\equiv 0\ \mathrm{or}\ \pi\ (\rm{mod}\ 2\pi).
\label{eq:invTR}
\end{equation}
We consider an implication of Supplementary Eq.~(\ref{eq:invTR}) for insulators and for 
nodal-line semimetals in the following discussion.


In a previous work \cite{PhysRevB.48.4442}, the 
relationship between the Zak phase and the surface polarization charge density 
$\sigma$ was
found. The surface polarization charge density, i.e. the surface normal component 
of the polarization vector, is given by 
\begin{equation}
\sigma=\sigma_{\mathrm{ion}}+\sigma_{e}
\label{eq:sigma}
\end{equation}
where  $\sigma_{\mathrm{ion}}$ is an ionic contribution 
from surface atoms, and $\sigma_{e}$ represents an electronic contribution
\begin{equation}
\sigma_e=\int\frac{d^2k_{\|}}{(2\pi)^2}\sigma_e({\bf k}_{\|}), \ \ 
\sigma_e({\bf k}_{\|})\equiv \frac{-e}{2\pi}\theta({\bf k}_{\|})\ \  (\mathrm{mod}\ e).
\end{equation}
If we regard the system at fixed ${\bf k}_{\|}$ to be a one-dimensional system, 
$\sigma_e({\bf k}_{\|})$ is an electronic surface charge density for the one-dimensional 
subssystem at ${\bf k}_{\|}$ \cite{PhysRevB.48.4442}.

We first consider insulators, assuming that there is no surface state that crosses the Fermi energy. 
Thus, $\sigma_e({\bf k}_{\|})$ does not have a jump as a function of ${\bf k}_{\|}$.
According to
Supplementary Eq.~(\ref{eq:invTR}), $\sigma_e({\bf k}_{\|})$ is therefore 
independent of $\mathbf{k}_{\|}$: 
\begin{equation}
\sigma_e({\bf k}_{\|})= N\frac{e}{2}, 
\label{eq:ins}
\end{equation}
where $N$ is an integer constant. Hence, the surface charge density is 
$\sigma_e=\frac{Ne}{2A_{\rm{surface}}}$ where $A_{\rm{surface}}$ is an area of the 
surface unit cell~\cite{PhysRevB.48.4442}. Although $N$ can be any integer, it is 
physically expected to vanish in almost all insulators, because nonzero $N$ 
corresponds to a large polarization, which leads to chemical or electronic instability. 
Thus $N=0$ is expected of stable electronic states; so far, no
insulator is known to have nonzero integer $N$, which means a huge surface polarization.

In materials with nodal lines which are the focus of the present work, 
the Zak phase jumps by $\pi$ at the nodal lines; therefore, there is always a region with $\theta({\bf k}_{\|})\equiv 0$ (mod $2\pi$) and
one with $\theta({\bf k}_{\|})\equiv \pi$ (mod $2\pi$). The latter region leads to an appreciable polarization. In nodal-line semimetals, 
the bulk electronic carriers and ions eventually screen the polarization, but  large deformation of the 
lattice structure and surface dipoles occur. We expect this to lead to large Rashba splitting if adatoms with large spin-orbit coupling  are present, as indicated in the main text.

\noindent
{\bf Supplementary Note 3. Surface termination and choice of the unit cell}

The surface
polarization charge density $\sigma_e$ is related to the polarization 
vector ${\bf P}$ by $\sigma=P_{\perp}\equiv{\bf P}\cdot{\bf n}$, where ${\bf n}$ is a
unit vector normal to the surface. 
Even when the direction of the surface plane is fixed, such as in (111) or (001), there are
possibilities for surface terminations. 
Moreover, there are various possible choices for the 
unit cell for a given surface termination.  

The dependence on the choice of unit cell is discussed in another study~\cite{PhysRevB.48.4442}. 
In summary, results of this work \cite{PhysRevB.48.4442} indicate that 
the polarization $\sigma(=P_{\perp})$ at a fixed surface termination is independent of the choice of the unit cell of the bulk. 
That is, whereas
$\sigma$
is independent of the choice of the unit cell at a fixed surface termination, the 
contributions of
$\sigma_e$ and $\sigma_{\mathrm{ion}}$ in Supplementary Eq.~(\ref{eq:sigma}) may depend on the 
unit cell choice. 

On the other hand, $\sigma$ 
generally changes with the surface termination. In the following discussion, 
we consider several cases of surface terminations 
for the (001) and (111) surfaces. For the calculations we always choose 
the unit cell in such a way that there are no additional `surface atoms', which 
are excess atoms that are not covered by translations of the unit cell 
~\cite{PhysRevB.48.4442}.
With such a choice of unit cell, we always have $\sigma_{{\rm ion}}=0$ and we only 
have 
 to consider the dependence of the electronic part $\sigma_{e}$. 

Thus, the unit cell is chosen accordingly in the following discussion of various surface terminations. 
Because the choice of unit cell corresponds to the unitary transformation of the Hamiltonian,
it affects the Zak phase in 
the following manner~\cite{PhysRevB.88.245126}. 
Suppose the unit structure consists of $N$ atoms at $\{
{\bf r}_1, \cdots,{\bf r}_N\}$. If the unit cell convention is changed 
to  $\{
{\bf r}_1+{\bf \epsilon}_1, \cdots,{\bf r}_N+{\bf \epsilon}_N\}$
where ${\bf \epsilon}_a$ ($a=1,\cdots,N$) are translation vectors of the
crystal, then the change in the Zak phase $\Delta\theta'({\bf k}_{\parallel})$ is
expressed as follows~\cite{PhysRevB.88.245126}:
\begin{equation}
\Delta\theta'({\bf k}_{\parallel})=-2\pi\sum_{a=1}^N
\epsilon_{a}^{\perp}\rho_a({\bf k}_{\parallel}),
\end{equation}
where $\epsilon_{a}^{\perp}$ is a surface-normal component of 
${\bf \epsilon}_{a}$, and 
\begin{equation}
\rho_a({\bf k}_{\parallel})\equiv\sum^{\rm occ.}_{m}\int_0^{b_{\perp}}\frac{dk_{\perp}}{2\pi}
\langle u_{n{\bf k}}|{\cal P}_a|u_{n{\bf k}}\rangle.
\end{equation}
Here ${\cal P}_a$ ($a=1,\cdots,N$) is a projection operator projecting onto the atom
$a$. 

In the present framework, 
${\bf \epsilon}_{a}$ is a translation vector, which 
is a linear combination of $\{{\bf a}_{1\|},{\bf a}_{2\|},{\bf a'} \}$ with 
integer coefficients. Among these primitive vectors, only ${\bf a'} $ has a
nonzero surface-normal component. Hence, $\epsilon_{a}^{\perp}$
is an integer multiple of $a'_{\perp}$.

\noindent
{\bf 3.1 (001) surface}

On the (001) surface, a $\sqrt{2}\times\sqrt{2}$ structure is formed when half of the surface atoms are
depleted. 
Therefore, we consider from the outset the unit cell for the 2D surface with an 
enlarged unit cell for the $\sqrt{2}\times\sqrt{2}$ structure. 
Let $a$ denote the lattice constant for the cubic unit cell of the fcc lattice. 
We take one of the surface atoms to be an origin, and the surface to be along the $xy$ plane.
The primitive vectors for the 2D surface can then be taken as:
\begin{equation}
{\bf a}_{1\|}=a(1,0,0),\ {\bf a}_{2\|}=a(0,1,0) 
\end{equation}
The other primitive vector is then given by ${\bf a}'=a(\frac{1}{2},0,\frac{1}{2}) $. 
The unit cell spanned by $\{{\bf a}_{1\|},{\bf a}_{2\|},{\bf a'} \}$ contains 
two atoms. Thus, we have
\begin{equation}
{\bf b}_{1}=\frac{2\pi}{a}(1,0,-1),\ \ 
{\bf b}_{2}=\frac{2\pi}{a}(0,1,0),\ \ 
{\bf b}_{\perp}=\frac{4\pi}{a}(0,0,1).
\end{equation}
Additionally,
\begin{equation}
{\bf b}_{1\|}=\frac{2\pi}{a}(1,0,0),\ \ 
{\bf b}_{2\|}=\frac{2\pi}{a}(0,1,0).
\end{equation}
For the perfect (001) surface on the $xy$ plane (Supplementary Figure~\ref{fccSM}a), the unit structure
consists of the two atoms at $(0,0,0)$ and ${\bf c}=a(-\frac{1}{2},\frac{1}{2},0)$.
Let us denote the two sublattices I and II, which belong to the points $0$ and 
${\bf c}$, respectively.
When the surface atoms at $a(m+\frac{1}{2},n+\frac{1}{2})$ ($m,n$: integer) become
depleted (Supplementary Figure~\ref{fccSM}b), the surface forms a $\sqrt{2}\times\sqrt{2}$ structure, 
and the unit structure
consists of the two atoms at $(0,0,0)$ and ${\bf c}+{\bf a}'=
a(0,\frac{1}{2},\frac{1}{2})$. Both choices of the 
unit structure are inversion-symmetric; therefore, 
the Zak phase in both cases is quantized as 0 or $\pi$ (mod $2\pi$).
Thus, the atom in sublattice II in the unit structure in
Supplementary Figure~\ref{fccSM}b is shifted from 
those in Supplementary Figure~\ref{fccSM}a
by ${\bf a}'$, 
and 
\begin{align}
&\Delta\theta'({\bf k}_{\parallel})=-2\pi
a'_{\perp}\rho_{\rm II}({\bf k}_{\parallel}),\\
&\rho_{\rm II}({\bf k}_{\parallel})\equiv\sum^{\rm occ.}_{m}\int_0^{b_{\perp}}\frac{dk_{\perp}}{2\pi}
\langle u_{n{\bf k}}|{\cal P}_{\rm II}|u_{n{\bf k}}\rangle.
\end{align}
Noting that the two sublattices are equivalent, we obtain
$\langle u_{n{\bf k}}|{\cal P}_{\rm a}|u_{n{\bf k}}\rangle=\frac{1}{2}\langle u_{n{\bf k}}|
u_{n{\bf k}}\rangle=\frac{1}{2}$ ($a={\rm I, II}$) and 
$\rho_{\rm II}({\bf k}_{\parallel})=\sum^{\rm occ.}_{m}\frac{1}{2}\frac{b_{\perp}}{2\pi}
=\frac{N_{\rm occ.}}{2}\frac{b_{\perp}}{2\pi}$, where $N_{\rm occ.}$ is the 
number of occupied bands. 
Thus,
\begin{equation}
\Delta\theta({\bf k}_{\parallel})=-
\frac{N_{\rm occ.}}{2}a'_{\perp}b_{\perp}=-\frac{N_{\rm occ.}}{2}2\pi
\end{equation}
Lastly we note that the unit cell is doubled from the original fcc unit cell;
therefore, $N_{\rm occ.}$
is even. Thus the Zak phase is unchanged, i.e. $
\Delta\theta({\bf k}_{\parallel})\equiv0\ \ ({\rm mod}\ 2\pi)$, in 
accordance with the {\it ab initio} calculation in the main text.
This invariance of the Zak phase is natural because it is a bulk quantity independent of the surface. 

\begin{figure}[ptb]
\centering 
\includegraphics[clip,width=0.45\textwidth ]{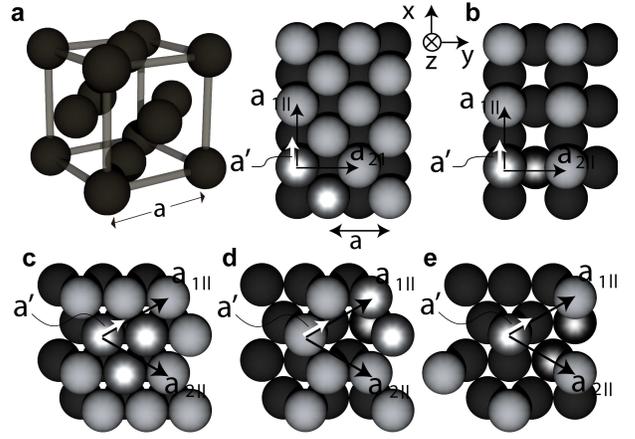} 
\caption{{\bf Surface structure and primitive vectors.}
{\bf a}, Crystal structure of  fcc Ca, Sr and Yb,
and that of the (001) surface (black circles) with surface atoms (grey circles).
{\bf b}, The same surface orientation but with one-half of the 
atoms per unit cell on the surface.
{\bf c}, Crystal structure of the (111) surface (black circles) with surface atoms (grey circles).
{\bf d}, {\bf  e}, The same surface orientation but with one-third and two-thirds of the 
atoms, respectively, per unit cell on the surface.
In {\bf a}-{\bf e}, the primitive vectors 
${\bf a}_{1\|}$, ${\bf a}_{2\|}$ along the surface are shown as black arrows, while
the other primitive vector 
${\bf a}'$ is shown as a white arrow.
Here, the choice of the unit structure is marked by the 
circles with gradation, i.e. in {\bf a} and {\bf b} the unit structure consists of two atoms, while 
in {\bf c}-{\bf e} it consists of three atoms.
}
\label{fccSM}
\end{figure} 

\noindent
{\bf 3.2 (111) surface}

When one-third or two-thirds of the surface atoms 
on the (111) surface are
depleted, a $\sqrt{3}\times\sqrt{3}$ structure is formed. 
Therefore, we consider from the outset the unit cell for the 2D surface with 
an enlarged unit cell for the $\sqrt{3}\times\sqrt{3}$ structure. 
We take one of the surface atoms to be an origin, and the surface to be along the $xy$ plane.
While the standard choice for the primitive vectors are 
$\tilde{{\bf a}}_{1\|}=\frac{a}{\sqrt{2}}(0,1,0),\ \tilde{{\bf a}}_{2\|}=\frac{a}{2\sqrt{2}}(-
\sqrt{3}, 1,0)$, 
the primitive vectors for the $\sqrt{3}\times\sqrt{3}$ structure can be 
\begin{equation}
{\bf a}_{1\|}=\frac{a}{2\sqrt{2}}(\sqrt{3},3,0) ,\ {\bf a}_{2\|}=\frac{a}{2\sqrt{2}}(-\sqrt{3},3,0) 
\end{equation}
The other primitive vector is then given by ${\bf a}'=\frac{a}{6\sqrt{2}}(\sqrt{3},3,2\sqrt{6}) $. 
The unit cell spanned by $\{{\bf a}_{1\|},{\bf a}_{2\|},{\bf a'} \}$ contains 
three atoms. Thus, we have
\begin{align}
&{\bf b}_{1}=\frac{2\pi}{3a}(\sqrt{6},\sqrt{2},-\sqrt{3}),\ \ 
{\bf b}_{2}=\frac{2\sqrt{2}\pi}{3a}(-\sqrt{3},1,0),\\
&{\bf b}_{\perp}=\frac{2\sqrt{3}\pi}{a}(0,0,1)
\end{align}
and 
\begin{equation}
{\bf b}_{1\|}=\frac{2\sqrt{2}\pi}{3a}(\sqrt{3},1,0),\ \ 
{\bf b}_{2\|}=\frac{2\sqrt{2}\pi}{3a}(-\sqrt{3},1,0).
\end{equation}
For the perfect (111) surface on the $xy$ plane (Supplementary Figure~\ref{fccSM}c), the unit structure
consists of the three atoms at $\{0,\tilde{{\bf a}}_{1\|},\tilde{{\bf a}}_{2\|}\}$.
Let us denote the three sublattices I, II and III which belongs to the points $0$, $\tilde{{\bf a}}_{1\|}$ and $\tilde{{\bf a}}_{2\|}$, respectively
When the one-third of the atoms become depleted (Supplementary Figure~\ref{fccSM}d), 
the unit structure
consists of the three atoms at $\{0,\tilde{{\bf a}}_{1\|}+{\bf a}',\tilde{{\bf a}}_{2\|}\}$.
Thus, the atom in sublattice II is shifted by ${\bf a}'$ in the new selected
of unit structure, 
and 
\begin{align}
&\Delta\theta({\bf k}_{\parallel})=-2\pi
a'_{\perp}\rho_{\rm II}({\bf k}_{\parallel}),\\
&\rho_{\rm II}({\bf k}_{\parallel})\equiv\sum^{\rm occ.}_{m}\int_0^{b_{\perp}}\frac{dk_{\perp}}{2\pi}
\langle u_{n{\bf k}}|{\cal P}_{\rm II}|u_{n{\bf k}}\rangle.
\end{align}
Noting that the three sublattices are equivalent, we obtain
$\langle u_{n{\bf k}}|{\cal P}_{\rm a}|u_{n{\bf k}}\rangle=\frac{1}{3}\langle u_{n{\bf k}}|
u_{n{\bf k}}\rangle=\frac{1}{3}$ ($a={\rm I, II, III}$), and 
$\rho_{\rm II}({\bf k}_{\parallel})=\sum^{\rm occ.}_{m}\frac{1}{3}\frac{b_{\perp}}{2\pi}
=\frac{N_{\rm occ.}}{3}\frac{b_{\perp}}{2\pi}$, where $N_{\rm occ.}$ is the 
number of occupied bands. 
Thus,
\begin{equation}
\Delta\theta({\bf k}_{\parallel})=-
\frac{N_{\rm occ.}}{3}a'_{\perp}b_{\perp}=-\frac{N_{\rm occ.}}{3}2\pi
\end{equation}
Lastly we note that the unit cell is tripled from the original fcc unit cell;
therefore, $N_{\rm occ.}$
is a integer multiple of three. Consequently, the Zak phase is unchanged: $
\Delta\theta({\bf k}_{\parallel})\equiv 0\ \ ({\rm mod}\ 2\pi)$.
This invariance of the Zak phase is natural because it is a bulk quantity independent of the surface.

When the two-thirds of the atoms are depleted (Supplementary Figure~\ref{fccSM}e), 
the unit structure
consists of the three atoms at $\{0,\tilde{{\bf a}}_{1\|}+{\bf a}',\tilde{{\bf a}}_{2\|}+{\bf a}'\}$.
Using a similar calculation we obtain
\begin{align}
&\Delta\theta'({\bf k}_{\parallel})=-2\pi(
a'_{\perp}\rho_{\rm II}({\bf k}_{\parallel})
+a'_{\perp}\rho_{\rm III}({\bf k}_{\parallel}))
,\\
&\rho_{\rm II}({\bf k}_{\parallel})=\rho_{\rm III}({\bf k}_{\parallel})=\frac{N_{\rm occ.}}{3}\frac{b_{\perp}}{2\pi}.
\end{align}
Therefore,
\begin{equation}
\Delta\theta({\bf k}_{\parallel})=-\frac{2N_{\rm occ.}}{3}2\pi\equiv 0\ \ ({\rm mod}\ 2\pi)
\end{equation}
and the Zak phase is unchanged, in 
accordance with the {\it ab initio} calculation in the main text.
We also note that in these three choices of the 
unit structure, the Zak phase is quantized as 0 or $\pi$ (mod $2\pi$).

\noindent
{\bf Supplementary Note 4. $\mathbb{Z}_2$ topology of nodal lines in alkaline-earth metals}

In a previous work~\cite{Fang15}, a $\mathbb{Z}_2$ topological number is defined for each nodal line 
in spinless systems with both inversion and time-reversal symmetries. 
If it is nontrivial, then the nodal line cannot vanish by itself after shrinking to a point. 
This $\mathbb{Z}_2$ topological number can be defined for each nodal line in 
Ca, when the spin-orbit interaction is neglected. This has been found to be 
trivial; the nodal lines around the L points
disappear upon addition of an artificial potential for the $4s$ orbital (see Supplementary Figures~\ref{z2SM}a-c).

Similar analysis of the nodal lines around 5 eV in Ag  (Supplementary Figure~\ref{z2SM}d) shows that it 
is also $\mathbb{Z}_2$-trivial  as defined by the above study~\cite{Fang15}. 
Supplementary Figures \ref{z2SM}e and f show the disappearance of the nodal line with the decrease in the on-site potential of the $5s$ orbital.

\begin{figure}[ptb]
\centering 
\includegraphics[clip,width=0.45\textwidth ]{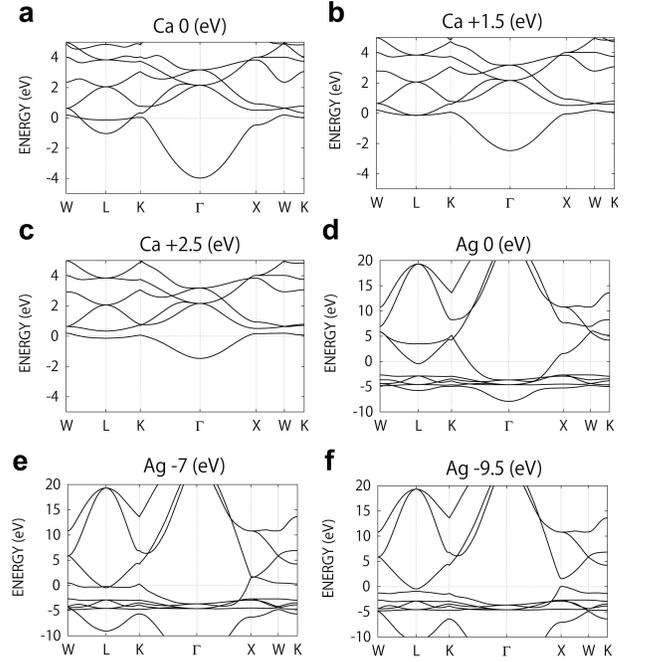} 
\caption{
{\bf Disappearance of nodal lines by adding the on-site potential.}
{\bf a}, Electronic band structure of Ca in the LDA. 
{\bf b}, {\bf c}, Electronic band structure of Ca,  depicting addition of $1.5$ and $2.5$ eV, respectively, to the on-site potential of the $s$ orbital.
{\bf d}, Electronic band structure of Ag in the LDA. 
{\bf e}, {\bf f}, Electronic band structure of Ag, depicting subtraction of $7$ and $9.5$ eV, 
respectively, from the on-site potential of the $s$ orbital.
}
\label{z2SM}
\end{figure} 

\noindent
{\bf Supplementary Note 5. Screening in nodal-line semimetals}

We have shown in the main text that when the nodal-line semimetal is regarded as a set of 
independent one-dimensional systems
for individual values of ${\bf k}_{\|}$, 
within a ${\bf k}$-space region of $\pi$ Zak phase, there is an appreciable polarization 
of $\pm  e/2$. Nevertheless, 
the polarization charges at the surface are eventually screened
since the entire system is a semimetal with carriers. 
In this section we consider screening of the surface polarization charges by carriers
in nodal-line semimetals. 
For simplicity, we consider the nodal-line semimetal with its nodal line 
being a circle in the $k_x$-$k_y$ plane with radius $k_0$, assuming the dispersion perpendicular to
the nodal line to be
linear with velocity $v_0$. The dispersion can then be
represented as $E=\pm \hbar v_0 \sqrt{(\sqrt{k_x^2+k_y^2}-k_0)^2+k_z^2}$. 
Thus, the density of states is $\nu(E)=C|E|$, with $C=\frac{k_0}{2\pi v_0^2\hbar^2}$ 
per unit volume.
In calcium there are four nodal lines; therefore, the constant $C$ 
is multiplied by the number of nodal lines $g(=4)$. 

The Poisson equation is 
\begin{equation}
\frac{d^2V}{dz^2}=\frac{e}{\varepsilon_0\varepsilon}\rho
\label{eq:Poisson}
\end{equation}
where $\rho(z)$ is the charge density, and $V$ is the potential energy for electrons~\cite{Zoellner}.
We set the $z$-axis 
normal to the surface of the semimetal, with $z=0$ representing the surface. 

We suppose that the polarization charge appears at the surface because of the 
presence of nodal lines, with polarization charge density $\sigma_{\rm s}$.
As we have shown in the main text, for example, 
the nodal line depletes electrons on the Ca surface within the area in ${\bf k}$ space 
surrounded by the nodal lines (shown as the shaded region in Supplementary Figure~3h), and the polarization charge is positive, that is, $\sigma_s>0$. Electron carriers are then induced near the surface 
because of  this positive surface charge, and $V(z)<0$ is expected for the region near the surface. 
The following equation relates the charge density $\rho$ to the potential $V$,  
\begin{equation}
\rho(z)=-en(z), \ n(z)= \int_0^{\infty}f_{\rm F}(E,z)\nu(E)dE, 
\label{eq:rhon}
\end{equation}
where $f_{\rm F}(E,z)=\frac{1}{e^{\beta(E-E_F+V(z))}+1}$ is the 
Fermi distribution function in the presence of potential $V(z)$. 
For simplicity we consider zero temperature and $E_F=0$ (i.e. at the nodal line). We thus have
\begin{equation}
n(z)=\frac{1}{2}CV(z)^2
\label{eq:n}
\end{equation}
From Eqs.~(\ref{eq:Poisson}) (\ref{eq:rhon}) and (\ref{eq:n}), we obtain
\begin{equation}
\frac{d^2V}{dz^2}=-\frac{e^2C}{2\varepsilon_0\varepsilon}V^2
\end{equation}
with boundary conditions $V(z=\infty)=0$, $V'(z=0)=\frac{e}{\varepsilon_0\varepsilon}\sigma_{\rm s}$.
The solution is 
\begin{equation}
V(z)=-\frac{e\lambda \sigma_{\rm s}}{2\varepsilon_0\varepsilon}\frac{1}{(1+z/\lambda )^2}, 
\end{equation}
where $\lambda =\left(\frac{24\varepsilon_0^2\varepsilon^2}{e^3C\sigma_{\rm s}}\right)^{1/3}
$ represents a screening length.
The charge distribution is
\begin{equation}
n(z)=\frac{1}{2}CV^2=\frac{C}{8}\left(\frac{e\lambda \sigma_{\rm s}}{\varepsilon_0\varepsilon}\right)^2\frac{1}{(1+z/\lambda )^4}
\end{equation} 
The spatial dependence of the potential $V(z)$ and 
electron density $n(z)$ are plotted
in Supplementary Figure~\ref{fig:screening}.
The total induced charge density is 
calculated as $\sigma_{\rm ind}=-e\int_0^{\infty}
ndz=-\sigma_s$. Therefore, the induced electronic distribution totally screens the positive polarization charge at the surface. 
Meanwhile, there remains a finite dipole moment, the density of which is calculated as follows:
\begin{equation}
-e\int_0^{\infty}
nzdz=-\frac{\sigma_{\rm s} \lambda}{2}
\end{equation}
Upon setting $\varepsilon=5$, $\sigma_{\rm s}\sim 0.243e/A_{\rm surface}$, 
$A_{\rm surface}=1.5\times 10^{-19}$m$^2$, $v_0\sim3\times 10^5 
{\rm m}\ {\rm s}^{-1}$, and $k_0\sim 0.24{\rm nm}^{-1}$ for rough estimates for calcium at 7.5 GPa,  
the screening length is estimated as $\lambda\sim 0.24{\rm nm}$, i.e. on the order of a lattice constant. The depth of the potential $V_0\equiv V(z=0)=-\frac{e\lambda \sigma_{\rm s}}{2\varepsilon_0\varepsilon}$ is approximately $-0.77$ eV. 
The dipole density per surface unit cell is 
$0.243e\cdot 0.24\mathrm{nm}/2=4.7\times 10^{-21}C\cdot \mathrm{nm}$, and 
the electric field at the surface is $-2V_0/(e\lambda)=6.4\mathrm{V}\ \mathrm{nm}^{-1}$.
\begin{figure}[ptb]
\centering 
\includegraphics[clip,width=0.25\textwidth ]{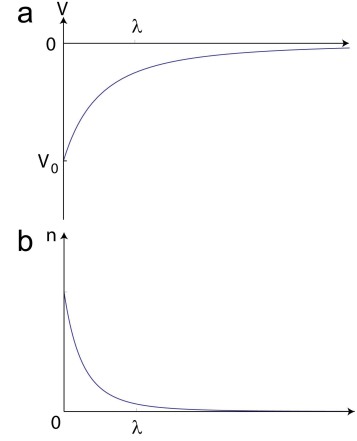} 
\caption{
{\bf Electronic potential and density due to screening of surface charges.}
{\bf a}, {\bf b}, Spatial dependence of the potential $V(z)$ and 
electron density, respectively.
}
\label{fig:screening}
\end{figure}  

\begin{figure}[ptb]
\centering 
\includegraphics[clip,width=8cm ]{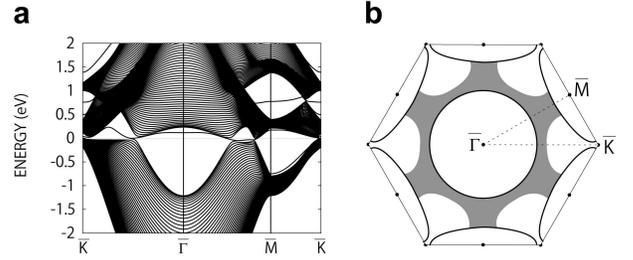} 
\caption{
{\bf Electronic relaxation on the surface of Ca.}
{\bf a}, Electronic band structure of Ca at 7.5 GPa for the (111) surface in the LDA, with the
lattice is  fixed. 
{\bf b}, The region in the surface Brillouin zone (grey); where the 
surface states descend below the Fermi energy by relaxation. The solid curves 
are projections of nodal lines.
}
\label{fig:surface-relaxed}
\end{figure}  

Thus far, we have studied screening by bulk carriers. We found that the 
dipoles are formed at the surface, and that the 
electronic potential is lower near the surface as shown in Supplementary Figure~\ref{fig:screening},  
This property affects surface states, if any, as discussed below.
Supplementary Figure \ref{fig:surface-relaxed} shows 
results from full self-consistent slab calculations with the lattice being fixed for 
the band structure of the 
Ca slab at 7.5GPa with 
(111) surfaces.
Comparing Supplementary Figure 3{\bf g} (without electronic relaxation)
and Supplementary Figure~\ref{fig:surface-relaxed}{\bf a} (with electronic relaxation), we see that 
surface states descend to the Fermi energy, which is within the grey region inSupplementary Figure~\ref{fig:surface-relaxed}{\bf b},
and that some of the surface states descend even below the  Fermi energy, becoming 
occupied (shown as the shaded region in Supplementary Figure~\ref{fig:surface-relaxed}{\bf b}). This lowering of surface states 
is attributed to the negative potential $V(z)$ near the surface.
Because the potential $V(z)$ is close to the surface, 
the surface states with shorter penetration depth are more affected by the 
potential $V(z)$. The maximum of the energy shift of the surface state
is expected to be $V_0$; it has been estimated to be around $-$0.77eV. This 
estimate is 
in good agreement with the energy shift of the surface states between 
Supplementary Figure~3{\bf g} (without electronic relaxation)
and Supplementary Figure~\ref{fig:surface-relaxed}{\bf a} (with electronic relaxation). To summarize, 
the 
bulk carriers partially screen the surface polarization charge due to the nodal lines, leaving behind dipoles at the surface. This induces an electronic potential 
which affects  surface states, if there are any surface states within the 
energy scale of the potential $V(z)$ at the surface.


\noindent
{\bf Supplementary References}

\end{document}